\documentclass[iop]{emulateapj}

\usepackage{amsmath}
\usepackage{color}

\newcommand{\secp}{\mbox{\rlap{.}$''$}} 

\newcommand{\angchem}{Angew. Chem. Int. Ed. Engl.}
\newcommand{\chemrev}{Chem. Rev.}
\newcommand{\cpl}{Chem. Phys. Lett.}
\newcommand{\cp}{Chem. Phys.}

\newcommand{\ijqc}{Int. J. Quantum Chem.}
\newcommand{\jms}{J. Mol. Spectr.}
\newcommand{\jmsttheochem}{J. Mol. Struct. (Theochem)}
\newcommand{\jpc}{J. Phys. Chem.}
\newcommand{\jpca}{J. Phys. Chem. A}
\newcommand{\nature}{Nature}
\newcommand{\psp}{Planet. Space Sci.}
\newcommand{\tca}{Theor. Chim. Acta}
\newcommand{\theochem}{J. Mol. Struct.: THEOCHEM}

\shorttitle{Discovery of interstellar CNCN}
\shortauthors{Ag\'undez et al.}

\begin{document}

\title{Discovery of interstellar isocyanogen (CNCN):\\ further evidence that dicyanopolyynes are abundant in space\thanks{Based on observations carried out with the IRAM 30m Telescope. IRAM is supported by INSU/CNRS (France), MPG (Germany) and IGN (Spain).}}

\author{M. Ag\'undez, N. Marcelino, \and J. Cernicharo}
\email{* marcelino.agundez@csic.es}
\affiliation{Instituto de F\'isica Fundamental, CSIC, C/ Serrano 123, 28006 Madrid, Spain}

\begin{abstract}

It is thought that dicyanopolyynes could be potentially abundant interstellar molecules, although their lack of dipole moment makes it impossible to detect them  through radioastronomical techniques. Recently, the simplest member of this chemical family, cyanogen (NCCN), was indirectly probed for the first time in interstellar space through the detection of its protonated form toward the dense clouds L483 and TMC-1. Here we present a second firm evidence of the presence of NCCN in interstellar space, namely the detection of the metastable and polar isomer isocyanogen (CNCN). This species has been identified in L483 and tentatively in TMC-1 by observing various rotational transitions in the $\lambda$~3 mm band with the IRAM 30m telescope. We derive beam-averaged column densities for CNCN of $1.6\times10^{12}$~cm$^{-2}$ in L483 and $9\times10^{11}$~cm$^{-2}$ in TMC-1, which imply fractional abundances relative to H$_2$ in the range $(5-9)\times10^{-11}$. While the presence of NCCN in interstellar clouds seems out of doubt owing to the detection of NCCNH$^+$ and CNCN, putting tight constraints on its abundance is still hampered by the poor knowledge of the chemistry that links NCCN with NCCNH$^+$ and especially with CNCN. We estimate that NCCN could be fairly abundant, in the range 10$^{-9}$-10$^{-7}$ relative to H$_2$, as other abundant nitriles like HCN and HC$_3$N.

\end{abstract}


\section{Introduction}

Interstellar chemistry is essentially organic in nature. Around three-fourths of the nearly 200 molecules detected to date in interstellar and circumstellar media contain at least one carbon atom. Among them, there are alcohols, aldehydes, acids, ethers, amines, but the most prevalent functional group is that of nitriles (e.g., \citealt{Agundez2013}). Indeed, the strong bond of the cyano group $-$C$\equiv$N is present in more than 30 interstellar molecules, although up to recently, no molecule containing two cyano groups had been observed in interstellar space \citep{Agundez2015b}.

A particular subclass of molecules with two cyano groups are dicyanopolyynes, which consist of a highly unsaturated linear skeleton of carbon atoms ended by a cyano group at each edge, i.e., N$\equiv$C$-$(C$\equiv$C)$_n-$C$\equiv$N. These molecules are very stable and it has been suggested that they could be abundant in interstellar and circumstellar clouds \citep{Kolos2000,Petrie2003}. The simplest member of this type of molecules, cyanogen (NCCN), is hypothesized to be a precursor of the widely observed cometary CN \citep{Fray2005} and it has been identified in the atmosphere of Titan \citep{Kunde1981}, where the larger homologue NC$_4$N has been long thought to be present as well \citep{Jolly2015}. A chemical cousin of cyanogen in which one N atom is substituted by a P atom, NCCP, has been tentatively identified in the carbon-rich envelope IRC\,+10216 \citep{Agundez2014}.

It is therefore reasonable to think that dicyanopolyynes can be abundant in molecular clouds. However, their detection is hampered by the fact that they are non polar and therefore cannot be observed through their rotational spectrum. To investigate the plausibility of this hypothesis it was proposed that the presence of NCCN in interstellar and circumstellar clouds can be indirectly probed through the observation of chemically related polar molecules, such as protonated cyanogen (NCCNH$^+$) or the metastable isomer CNCN \citep{Petrie2003}. A few years ago, we succeeded in detecting protonated cyanogen in the cold dark clouds L483 and TMC-1 \citep{Agundez2015b}. From the observed abundances of NCCNH$^+$ it was inferred that indeed NCCN should be as abundant as the nitriles HCN and HC$_3$N. In this Letter, we present the identification of isocyanogen (CNCN), a metastable isomer of NCCN, in L483 and a tentative detection toward TMC-1. This result brings further support to the hypothesis that dicyanopolyynes are abundant molecules in interstellar space.

\section{Observations}

The observations of L483 were carried out with the IRAM 30m telescope in the course of a $\lambda$~3 mm molecular line survey. Details on these observations are given in  \citet{Agundez2018} and \citet{Marcelino2018}, while a thorough description of the complete survey will be presented elsewhere (Ag\'undez et al., in preparation). Briefly, observations were performed in several sessions from August 2016 to April 2018, with the telescope pointed toward the position of the infrared source IRAS\,18148$-$0440. We used the frequency-switching technique, with the receiver EMIR E090 connected to the FTS backend with a spectral resolution of 50 kHz.

The observations of TMC-1 were also taken with the IRAM 30m telescope as part of a $\lambda$~3 mm spectral scan at the position of the cyanopolyyne peak \citep{Marcelino2005,Marcelino2007,Marcelino2009}. A large fraction of the data were observed in February 2012 (see details in \citealt{Cernicharo2012}).

\section{Results}

\begin{figure}
\centering
\includegraphics[angle=0,width=\columnwidth]{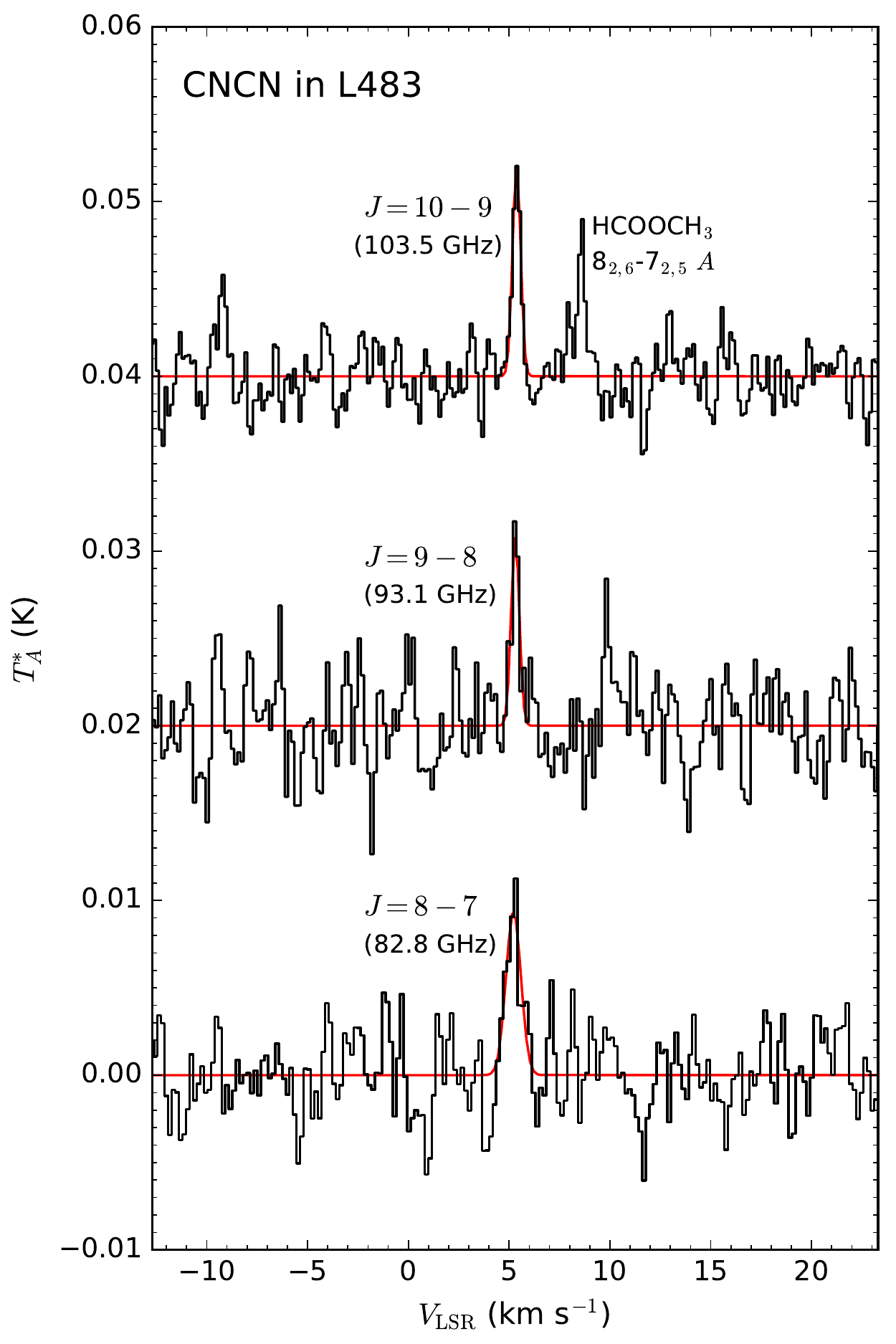}
\caption{Lines of CNCN observed toward L483 with the IRAM 30m telescope. Lines are detected at 8-14$\sigma$ confidence levels. The $J=11-10$ line at 113.8 GHz is not detected due to the more limited sensitivity of the data at these frequencies.} \label{fig:lines_l483}
\end{figure}

\begin{table}
\caption{Observed line parameters of CNCN in L483} \label{table:lines}
\centering
\begin{tabular}{lrclc}
\hline \hline
\multicolumn{1}{c}{Transition} & \multicolumn{1}{c}{Frequency} & \multicolumn{1}{c}{$V_{\rm LSR}$} & \multicolumn{1}{c}{$\Delta v$}      & \multicolumn{1}{c}{$\int T_A^* dv$} \\
                                                & \multicolumn{1}{c}{(MHz)}                 & \multicolumn{1}{c}{(km s$^{-1}$)}    & \multicolumn{1}{c}{(km s$^{-1}$)} & \multicolumn{1}{c}{(K km s$^{-1}$)} \\
\hline
$J=8-7$     &   82784.692 & +5.21(7) & 0.87(14) & 0.0087(13) \\ 
$J=9-8$     &   93132.326 & +5.30(4) & 0.47(14) & 0.0052(4)   \\ 
$J=10-9$   & 103479.802 & +5.37(3) & 0.47(6)   & 0.0061(6)   \\ 
\hline
\end{tabular}
\tablecomments{Numbers in parentheses are 1$\sigma$ uncertainties in units of the last digits obtained from Gaussian fits.}
\end{table}

The analysis of the $\lambda$~3 mm IRAM 30m data of L483 has already resulted in the detection of several new interstellar molecules: HCCO \citep{Agundez2015a}, NCCNH$^+$ \citep{Agundez2015b}, NS$^+$ \citep{Cernicharo2018}, HCS and HSC \citep{Agundez2018}, as well as NCO and the confirmation of H$_2$NCO$^+$ \citep{Marcelino2018}. A complete analysis of the $\lambda$~3 mm molecular line survey will be presented in a separate article (Ag\'undez et al., in preparation). Here we focus on three emission lines which can be unambiguously assigned to the $J=8-7$, $J=9-8$, and $J=10-9$ rotational transitions of CNCN, lying at 82.8 GHz, 93.1 GHz, and 103.5 GHz (see Fig.~\ref{fig:lines_l483} and Table~\ref{table:lines}).

Isocyanogen (CNCN), a metastable isomer lying 1.06 eV above cyanogen (NCCN), is a linear molecule with a $^1\Sigma^+$ electronic ground state \citep{Botschwina1990,Ding2000,Chaudhuri2006}. Its rotational spectrum has been extensively characterized in the laboratory from 70 to 340 GHz \citep{Winnewisser1992}. The hyperfine structure due to the quadrupole of the two $^{14}$N nuclei has been measured for the $J=1-0$ and $J=2-1$ lines, where the various components are separated by a few MHz \citep{Gerry1990}. However, the higher$-J$ lines observed here have very small line splittings and cannot be spectrally resolved (see JPL catalog\footnote{\texttt{https://spec.jpl.nasa.gov/}}; \citealt{Pickett1998}). Isocyanogen is a moderately polar molecule, with a measured permanent electric dipole moment of 0.7074 $\pm$ 0.0052 D \citep{Gerry1990}. The species is included in the MADEX code\footnote{\texttt{https://nanocosmos.iff.csic.es/?page\_id=1619}} \citep{Cernicharo2012:madex}.

The spectra of L483 at 82.8 and 93.1 GHz have $T_A^*$ rms noise levels of 2.2-2.3 mK per 50 kHz channel, which translates to detection confidence levels of 8-9$\sigma$ for the $J=8-7$ and $J=9-8$ lines of CNCN. We note that the Gaussian fit to the $J=8-7$ line results in a too large linewidth (0.87 km s$^{-1}$ vs. the value of 0.47 km s$^{-1}$ obtained for the two other lines; see Table~\ref{table:lines}) which is probably a consequence of the presence of positive channels around the line which are within the noise but are taken as part of the line by the fitting algorithm. The spectrum at 103.5 GHz is more sensitive, with a $T_A^*$ rms of 1.6 mK per 50 kHz channel, and the $J=10-9$ line is detected at a higher confidence level of 14$\sigma$. This spectrum does also clearly show a weak line arising from methyl formate (see Fig.~\ref{fig:lines_l483}). The line survey indeed shows many more lines of this complex organic molecule, which has been recently detected in L483 using ALMA \citep{Oya2017}. There is another rotational transition of CNCN, the $J=11-10$ lying at 113.8 GHz, which falls within the frequency range covered by our line survey, although in this spectral region the data is less sensitive (the $T_A^*$ rms noise level is $\sim$5 mK per 50 kHz channel) and the line is not clearly detected. With three lines well detected, we consider that the identification of CNCN in L483 is completely secure.

\begin{figure}
\centering
\includegraphics[angle=0,width=\columnwidth]{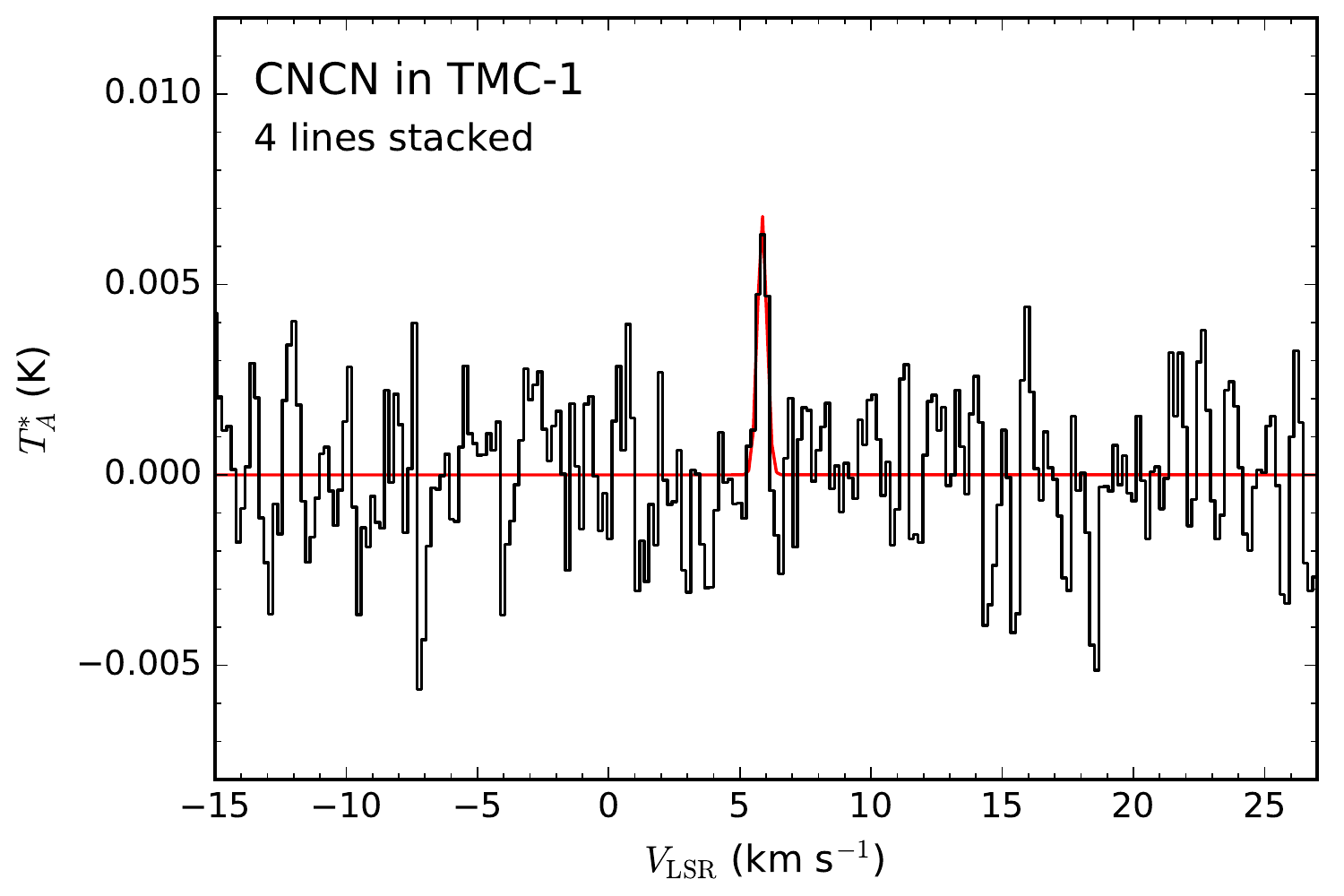}
\caption{IRAM 30m spectrum obtained toward TMC-1 by stacking four lines, the $J=8-7$ through $J=11-10$, of CNCN. The line is detected at a 5-6$\sigma$ confidence level.} \label{fig:lines_tmc1}
\end{figure}

We have also searched for CNCN in our $\lambda$~3 mm data of TMC-1 obtained with the IRAM 30m telescope. The data covers the four rotational transitions, $J=8-7$ through $J=11-10$. None of them are detected at a sufficient confidence level, but after stacking the spectra around the four lines, an emission line with a peak intensity of $T_A^*\sim7$ mK appears centered at $V_{\rm LSR}$ of 5.85 $\pm$ 0.05 km s$^{-1}$, in agreement with the systemic velocity of TMC-1 measured in other molecular lines (e.g., \citealt{Agundez2015b}). The line is detected at a 5-6$\sigma$ confidence level. However, taking into account that it comes from a line stacking of four lines, we consider the detection as tentative. More sensitive observations are needed to detect with a sufficient signal-to-noise ratio the individual lines.


From the intensities of the three CNCN lines detected in L483 and assuming a rotational temperature of 10 K, which is the gas kinetic temperature in this dense core as derived from NH$_3$ \citep{Anglada1997}, we find a beam-averaged column density for CNCN of $1.6\times10^{12}$ cm$^{-2}$. In TMC-1 we estimate $N$(CNCN) = $9\times10^{11}$ cm$^{-2}$, assuming a rotational temperature of 10 K \citep{Pratap1997}. Adopting H$_2$ column densities of $3\times10^{22}$ cm$^{-2}$ for L483 \citep{Tafalla2000} and $1\times10^{22}$ cm$^{-2}$ for TMC-1 \citep{Cernicharo1987}, the abundance of CNCN relative to H$_2$ is then $5\times10^{-11}$ in L483 and $9\times10^{-11}$ in TMC-1. The metastable isomer CNCN is therefore significantly more abundant than protonated cyanogen, with column density ratios CNCN/NCCNH$^+$ of 40 and 10 in L483 and TMC-1, respectively.

The dense cores TMC-1 and L483 are rich in carbon chain molecules (e.g., \citealt{Kawaguchi1995}; Ag\'undez et al., in preparation), and this common feature may be at the origin of the presence of cyanogen in both clouds. The two sources are however different in that TMC-1 is a starless core which displays narrow lines \citep{Soma2018}, while L483 host a Class\,0 protostar and some molecules show wide lines whose emission arises from an outflow (e.g., H$_2$CO; \citealt{Tafalla2000}) or from a compact hot corino around the protostar, as revealed by recent ALMA observations (e.g., HNCO; \citealt{Oya2017}). The narrow line profiles of CNCN in L483 point to emission from the ambient cloud. We estimate a $3\sigma$ upper limit to the abundance of CNCN in the hot corino of $(5-7)\times10^{-9}$ relative to H$_2$, adopting a source diameter of 0\secp5, a line width of 10 km s$^{-1}$, a rotational temperature of 70-130 K, and $N$(H$_2$) = $6.5\times10^{23}$ cm$^{-2}$ (see \citealt{Oya2017}).

\section{Discussion}

The discovery of CNCN opens a new venue, in addition to that provided by NCCNH$^+$, to probe the presence of the stable and likely abundant, although non polar, cyanogen molecule (NCCN). However, in order to use these species as markers of cyanogen it is necessary to understand the underlying chemical relationships between them.

The chemistry of NCCNH$^+$ and its chemical connection with NCCN was studied by \citet{Petrie2003} and more recently by \citet{Agundez2015b}. In both studies the NCCNH$^+$/NCCN abundance ratio predicted from a cold dark cloud chemical model is of the order of 10$^{-4}$. Therefore, the observed abundances of NCCNH$^+$ put the abundance of NCCN in dark clouds at the level of $(1-10)\times10^{-8}$ relative to H$_2$ \citep{Agundez2015b}. We note however that in this latter study, the observed abundance ratios between the protonated and non protonated forms of the related cyanides HCN (and HNC) and HC$_3$N are underestimated by the chemical model by roughly a factor of ten. If the same holds for NCCN, its abundance would need to be revised downward to $(1-10)\times10^{-9}$ relative to H$_2$.

In the case of the isomer CNCN, little is known about its chemistry and thus it is difficult to relate the observed abundance to that of NCCN. However, as noted by \citet{Petrie2003}, this pair of isomers share structural characteristics (namely, a $-$CNC vs. a $-$CCN subunit) with HCCNC and HCCCN, both of which are observed in cold dark clouds. Concretely, in L483 and TMC-1 the HCCNC/HCCCN column density ratio is around 50 (details for L483 will follow in Ag\'undez et al., in preparation; see, e.g., \citealt{Agundez2013} for TMC-1). If the CNCN/NCCN abundance ratio is similar, the observed abundances of CNCN in L483 and TMC-1 would imply a NCCN abundance of $(2.5-4.5)\times10^{-9}$ relative to H$_2$. It is worth noting that these values are in the range of abundances inferred above from NCCNH$^+$ adopting a NCCNH$^+$/NCCN abundance ratio of 10$^{-3}$, instead of 10$^{-4}$.

In any case, it is clear that a better understanding of the chemistry of NCCN, CNCN, and NCCNH$^+$ in interstellar clouds is necessary to draw more robust conclusions. It is therefore worth to revisit what is known and in which directions further research is needed.

It is well established that the reaction between CN and HCN, which is exothermic to yield cyanogen, has an important activation barrier that makes it too slow at low temperatures \citep{Yang1992}. Therefore, the most plausible formation route of NCCN in cold clouds is the reaction
\begin{subequations}
\begin{eqnarray}
\rm CN + HNC & \rightarrow & \rm NCCN + H, \label{reac:cn+hnc=nccn} \\
                        & \rightarrow & \rm CNCN + H,
\end{eqnarray} \label{reac:cn+hnc}
\end{subequations}
which involves two abundant reactants and is barrierless to produce NCCN, although not when forming CNCN \citep{Petrie2003}. It would nevertheless be interesting to revisit this reaction through experiments or dynamical calculations to quantify the kinetics of the various possible channels. It is difficult to envisage other feasible (rapid and involving abundant reactants) NCCN-forming reactions. Another possible route is the reaction between N and HCCN, for which \citet{Loison2015} estimate a moderate rate coefficient of a few times 10$^{-11}$ cm$^3$ s$^{-1}$, although this reaction would be less efficient than reaction~(\ref{reac:cn+hnc=nccn}) owing to the low abundance of HCCN. Reactions of N$_2$ with radicals are unlikely to be exothermic owing to the strong triple bond of molecular nitrogen. For example, the reaction between N$_2$ and C$_2$H to produce NCCN is endothermic by $\sim$50 kJ mol$^{-1}$.

\begin{figure}
\centering
\includegraphics[angle=0,width=\columnwidth]{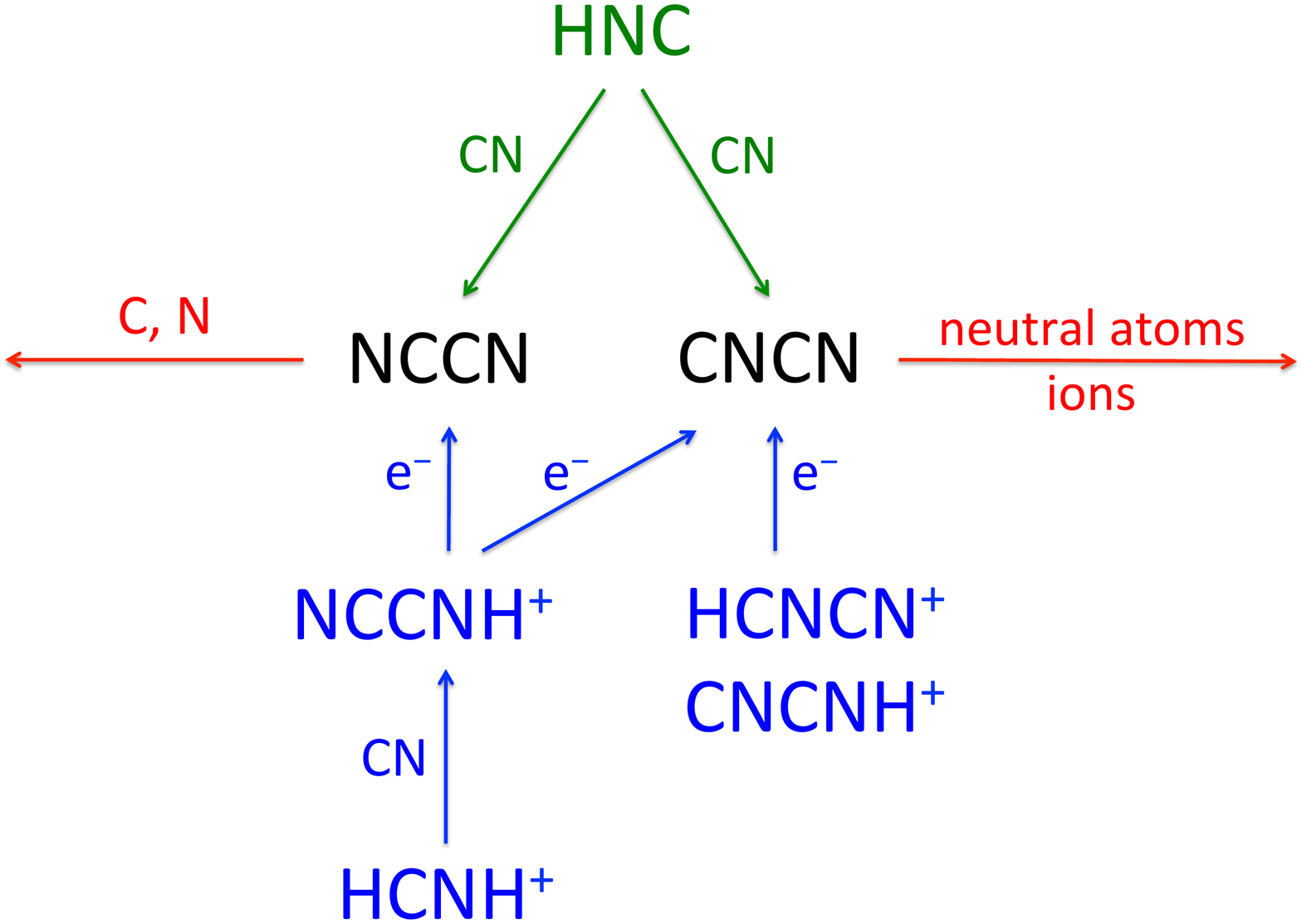}
\caption{Main formation (neutral in green and ionic in blue) and destruction (in red) pathways for NCCN and CNCN. The low-temperature kinetics of these reactions is very poorly constrained.} \label{fig:scheme}
\end{figure}

If neutral-neutral gas-phase routes are not efficient enough, routes involving ions may do the work. The most obvious is
\begin{equation}
\rm NCCNH^+ + e^- \rightarrow products, \label{reac:nccnh++e-}
\end{equation}
which could reasonably lead to NCCN + H or CNCN + H, although neither the rate coefficient nor the product distribution have been measured. A similar mechanism involving the ion HCNH$^+$ is thought to form HCN and HNC in dark clouds (e.g., \citealt{Agundez2013}). Note that if NCCNH$^+$ is the precursor of NCCN, and not the reverse, as occurs if NCCN is mainly formed by neutral-neutral routes like reaction~(\ref{reac:cn+hnc=nccn}), then the NCCNH$^+$/NCCN abundance ratio will no longer be of the order of 10$^{-4}$, but significantly higher. The question then would be how can NCCNH$^+$ be efficiently formed, typically from smaller fragments. One possible candidate is
\begin{equation}
\rm HCNH^+ + CN \rightarrow NCCNH^+ + H, \label{reac:hcnh++cn}
\end{equation}
which is a potentially important source of NCCNH$^+$ because the reactants are relatively abundant. Reaction~(\ref{reac:hcnh++cn}) is nearly thermoneutral and thus it is unclear whether or not it is rapid at low temperatures. The reaction between HCN$^+$ and HCN is exothermic in the channel yielding NCCNH$^+$. However, the main channel appears to be HCNH$^+$ + CN \citep{Huntress1969} and besides, the reactant HCN$^+$ should be not very abundant in dark clouds. Another potentially interesting reactions forming NCCNH$^+$ are C$_2$N$^+$ + NH$_2$ and C$_2$NH$^+$ + NH, although the reactants are probably not abundant enough in dark clouds. The isomer CNCN can be formed in the dissociative recombination of NCCNH$^+$ with electrons if some important rearrangement takes place. A study of the product distribution of reaction~(\ref{reac:nccnh++e-}) would allow to validate or refute this possibility. If NCCNH$^+$ is not a viable precursor for CNCN, then protonated isocyanogen (either HCNCN$^+$ or CNCNH$^+$, both having similar energies; \citealt{Petrie1998,Ding2001}) could be the main precursor. However, protonated isocyanogen is 60-65 kJ mol$^{-1}$ less stable than protonated cyanogen \citep{Petrie1998} and thus forming it would be more difficult. For example, while a priori reaction~(\ref{reac:hcnh++cn}) is a viable route to NCCNH$^+$, the channel leading to HCNCN$^+$ or CNCNH$^+$ is probably endothermic.

A synthetic route to NCCN and CNCN alternative to the gas phase could take place on dust grains by recombination of two CN radicals. It is however uncertain whether the abundance of CN on grain surfaces is high enough. The CN radical, as other nitrogen-bearing species, tend to show little depletion from the gas phase in dense cores \citep{Hily-Blant2008}. It is also not clear whether such reaction may have some barrier and ultimately how effective would be the desorption of these species to the gas phase.

Whatever the mechanism of formation of NCCN, it is likely to have a long life time in cold dark clouds because it is not a very reactive species, although the same may not be true for CNCN. Cyanogen shows a moderate reactivity with ions \citep{McEwan1995} and it does not react at low temperature with H or O atoms (see NIST database\footnote{\texttt{https://kinetics.nist.gov/kinetics/}}). It however reacts with C and N atoms
\begin{equation}
\rm NCCN + C \rightarrow CN + C_2N, \label{reac:nccn+c}
\end{equation}
\begin{equation}
\rm NCCN + N \rightarrow N_2 + C_2N, \label{reac:nccn+n}
\end{equation}
with rate coefficients of a few times 10$^{-11}$ cm$^3$ s$^{-1}$ at room temperature \citep{Whyte1983,Safrany1968}. However, taking into account that reactions~(\ref{reac:nccn+c}) and (\ref{reac:nccn+n}) are thought to be the main depletion mechanism of NCCN \citep{Agundez2015b}, it would be interesting to study their low-temperature kinetics.

\section{Prospects}

There are various directions to advance in our understanding of the presence and abundance of cyanogen and dicyanopolyynes in interstellar space. A first one would be to study reactions~(\ref{reac:cn+hnc}-\ref{reac:nccn+n}) either experimentally or theoretically (see Fig.~\ref{fig:scheme}) to better understand the chemical relationship of NCCNH$^+$ and CNCN with the likely abundant but unobservable NCCN.

The astronomical search for other markers of cyanogen is also an interesting possibility. For example, apart from CNCN, there are other two linear metastable isomers of NCCN. These are di-isocyanogen (CNNC) and diaza-dicarbon (NNCC), isomeric forms that lie higher in energy, specifically at 3.16 and 3.90 eV, respectively, above NCCN \citep{Ding2000,Chaudhuri2006}. While CNNC is non polar, NNCC is calculated to have a large dipole moment of 3.35 D \citep{Ding2000}, and it is thus a candidate for interstellar detection. This isomer however seems elusive to laboratory detection \citep{Maier1992} and therefore its rotational spectrum remains unknown. An astronomical search for the protonated form of CNCN, either HCNCN$^+$ or CNCNH$^+$ \citep{Petrie1998,Ding2001}, would also provide valuable information on the chemical origin of CNCN, but the rotational spectrum has not been measured.

If cyanogen is really an abundant species in interstellar clouds, it is conceivable that larger members of the series of dicyanopolyynes are present as well. The next member is dicyanoacetylene (NC$_4$N), which, as NCCN, is non polar and thus indirect strategies to probe it are necessary. Dicyanoacetylene can be efficiently formed through the reactions CN + HC$_3$N or C$_3$N + HCN/HNC, which are barrierless according to calculations by \citet{Petrie2004}. An astronomical search for the protonated form of NC$_4$N is not yet feasible as there is little spectroscopic information on this ion. However, the metastable isomer NC$_3$NC, which lies 1.12 eV above NC$_4$N has a moderate calculated dipole moment of 1.04-1.096 D \citep{Horn1994,Lee1998}, and its rotational spectrum has been measured in the laboratory \citep{Huckauf1999}. It is not present in our $\lambda$~3 mm data of L483 and TMC-1, although this is probably not the best wavelength range to search for such a heavy molecule in cold sources. A search at longer wavelengths may probe more fruitful. There are other two metastable isomers, but one (CNC$_2$NC) is non polar while for the other (NCNC$_3$), which lies 2.60 eV above NC$_4$N \citep{Jiang2004}, spectroscopic data is lacking.

A better understanding of the chemistry of NCCN in interstellar space would benefit from the search for the already detected species NCCNH$^+$ and CNCN in other sources, ranging from prestellar cores to star-forming regions. Getting access to the spatial distribution of these species in the targeted clouds will also help, although lines are weak and obtaining emission maps will be difficult, especially if NCCN and their markers NCCNH$^+$ and CNCN have an extended distribution over the ambient cloud, which is the most likely possibility. Finally, a direct search for NCCN at infrared wavelengths with telescopes like SOFIA or JWST may probe fruitful toward strong infrared sources.
 
\section{Conclusions}

We have reported on the first identification in space of isocyanogen (CNCN), a metastable isomer of cyanogen. This species has been observed in the dense core L483 and tentatively in TMC-1, where we previously reported the detection of protonated cyanogen (NCCNH$^+$). We derive fractional abundances of several times 10$^{-11}$ relative to H$_2$ for CNCN. The detection of CNCN together with that of NCCNH$^+$ in dark clouds supports the idea that NCCN and larger dicyanopolyynes, all of which are non polar and thus cannot be directly observed through their rotational spectrum, are abundant in molecular clouds. We estimate that the abundance of NCCN may be in the range 10$^{-9}$-10$^{-7}$ relative to H$_2$. A better knowledge of the chemistry of NCCN and dicyanopolyynes in space will benefit from experimental and/or theoretical studies of some key reactions together with sensitive astronomical observations of NCCNH$^+$, CNCN, and related species not yet identified in space.

\acknowledgments

We thank the IRAM 30m staff for their help during the observations and the anonymous referee for useful comments. We acknowledge funding support from the European Research Council (ERC Grant 610256: NANOCOSMOS) and from Spanish MINECO through grant AYA2016-75066-C2-1-P. M.A. also acknowledges funding support from the Ram\'on y Cajal programme of Spanish MINECO (RyC-2014-16277).

\end{document}